\documentclass{iau}
\usepackage{graphicx}

\title[IAUS302.~~NSMAXG: Neutron star spectral model]
{NSMAXG: A new magnetic neutron star spectral model in XSPEC}

\author[W.~C.~G.~Ho]{Wynn C.~G.~ Ho$^1$}

\affiliation{$^1$Mathematical Sciences and STAG,
University of Southampton, Southampton, SO17 1BJ, UK \\
email: {\tt wynnho@slac.stanford.edu}}

\pubyear{2013}
\volume{302}  
\pagerange{1--4}
\jname{Magnetic Fields Throughout Stellar Evolution}
\editors{M. Jardine, P. Petit \& H.C. Spruit, eds.}
\begin{document}

\maketitle

\begin{abstract}
The excellent sensitivity of X-ray telescopes, such as Chandra and XMM-Newton,
is ideal for the study of cooling neutron stars, which can emit at these
energies. In order to exploit the wealth of information contained in the
high quality data, a thorough knowledge of the radiative properties of
neutron star atmospheres is necessary.
A key factor affecting photon emission is magnetic fields, and
neutron stars are known to have strong surface magnetic fields.
Here I briefly describe our latest work on constructing magnetic
($B\ge 10^{10}\mbox{ G}$) atmosphere models of neutron stars
and the NSMAXG implementation of these models in XSPEC.
Our results allow for more robust extractions of neutron star parameters
from observations.
\end{abstract}

Thermal X-ray radiation has been detected from many radio pulsars and
radio-quiet neutron stars.
Thermal emission can provide invaluable information on the physical
properties and evolution of neutron stars,
such as the mass $M$, radius $R$, and surface temperature $T$,
which in turn depend on poorly constrained physics of the deep interior,
such as the nuclear equation of state and quark and superfluid and
superconducting properties at supranuclear densities.
In addition, neutron stars are known to possess strong magnetic fields:
from $B\approx 10^{8}-10^{9}\mbox{ G}$ in the case of millisecond pulsars
to $B\approx 10^{10}-10^{13}\mbox{ G}$ in the case of normal pulsars and
even $B\gtrsim 10^{14}\mbox{ G}$ in magnetars
(see \cite[Zavlin 2009]{zavlin09}; \cite[Kaspi 2010]{kaspi10};
\cite[Harding 2013]{harding13}, for reviews).

The observed thermal radiation originates in a thin atmospheric layer
(with scale height $\sim 1\mbox{ cm}$) that covers the stellar surface.
Atmosphere properties, such as magnetic field, chemical composition, and
radiative opacities, directly determine the characteristics of the
observed spectrum.
While the surface composition of neutron stars is generally unknown, a great
simplification arises due to the efficient gravitational separation of light
and heavy elements (\cite[Alcock \& Illarionov 1980]{alcockillarionov80};
\cite[Hameury et al. 1983]{hameuryetal83}).
A pure hydrogen atmosphere is expected even if a small amount of
accretion occurs after neutron star formation; the total mass of
hydrogen needed to form an optically thick atmosphere can be less than
$\sim 10^{16}\mbox{ g}$.
Alternatively, a helium or carbon atmosphere may be possible as a result
of nuclear burning on the neutron star surface
(\cite[Chang \& Bildsten 2003; Chang et al. 2010]{changbildsten03,changetal10}).
Finally, a heavy-element atmosphere may exist if no accretion takes place
or if all the accreted matter is consumed by nuclear reactions.

Steady progress has been made in modeling neutron star atmospheres (see
\cite[Zavlin 2009]{zavlin09}, for more detailed discussion and references).
Since the neutron star surface emission is thermal in nature, it has been
modeled at the lowest approximation with a blackbody spectrum.  Early works
on more realistic spectra assumed emission from unmagnetized light-element
atmospheres, and the resultant spectra exhibit distinctive hardening
relative to a blackbody.
The inclusion of magnetic fields has many important effects.
For example, the presence of a magnetic field causes emission to be
anisotropic and polarized
(see \cite[M\'{e}sz\'{a}ros 1992]{meszaros92}, for review).
At $B>e^3m_{\mathrm{e}}^2c/\hbar^3=2.35\times 10^{9}\mbox{ G}$,
the binding energy of atoms, molecules, and other bound states increases
significantly, and abundances can be appreciable in the atmosphere of
neutron stars (see \cite[Lai 2001]{lai01}, for review).
When field strengths approach and exceed the quantum electrodynamics field
$B_{\mathrm{QED}}=m_{\mathrm{e}}^2c^3/e\hbar=4.41\times 10^{13}\mbox{ G}$,
vacuum polarization effects, such as switching between photon polarization
modes, become relevant (\cite[Lai \& Ho 2002]{laiho02};
\cite[van Adelsberg \& Lai 2006]{vanadelsberglai06}),
and the atmosphere may even cease to exist as plasma condenses onto the surface
(\cite[Medin \& Lai 2007; Potekhin et al. 2012]{medinlai07,potekhinetal12}).
Most calculations of magnetic neutron star atmospheres focus on a fully
ionized hydrogen plasma.
Only relatively recently have self-consistent atmosphere models using the
latest equation of state and opacity results for strongly magnetized and
{\it partially ionized} hydrogen and mid-Z elements been constructed
(\cite[Potekhin et al. 2004]{potekhinetal04}; \cite[Mori \& Ho 2007]{moriho07}).

In previous work, we implemented into XSPEC (\cite[Arnaud 1996]{arnaud96})
our theoretical neutron star magnetic atmosphere X-ray spectra,
under the model name NSMAX (\cite[Ho et al. 2008]{hoetal08}).
These atmosphere spectra are obtained using the partially ionized results of
\cite[Potekhin et al. (2004)]{potekhinetal04} and
\cite[Mori \& Ho (2007)]{moriho07}.
Two sets of models are provided: One set with a single surface ${\mathbf{B}}$
and $T_{\mathrm{eff}}$ and a second set which is constructed with
${\mathbf{B}}$ and $T_{\mathrm{eff}}$ varying across the surface according
to a magnetic dipole geometry.
Magnetic fields and effective temperatures of the models span the range
$B=10^{12}-3\times 10^{13}\mbox{ G}$ and
$\log T_{\mathrm{eff}}\mbox{(K)}\approx 5.5-6.7$, respectively.
Note that other neutron star atmosphere spectra in XSPEC are either
non-magnetic (NSAGRAV: \cite[Zavlin et al. 1996]{zavlinetal96};
NSSPEC: \cite[G\"{a}nsicke et al. 2002]{gansickeetal02};
NSATMOS: \cite[McClintock et al. 2004]{mcclintocketal04};
\cite[Heinke et al. 2006]{heinkeetal06}) or magnetic but fully ionized
hydrogen (NSA: \cite[Pavlov et al. 1995]{pavlovetal95});
the last at only two fields: $B=10^{12}$ and $10^{13}\mbox{ G}$.
We also note the open source non-magnetic model McPHAC
(\cite[Haakonsen et al. 2012]{haakonsenetal12}).

We recently implemented into XSPEC a new set of neutron star magnetic
atmosphere X-ray spectra, under the model name NSMAXG, which replaces NSMAX.
The new model is nearly identical to the old model but with two important
differences.
The first difference is the inclusion of atmosphere spectra for weaker
magnetic fields ($B=10^{10}-10^{11}\mbox{ G}$).
These spectra are constructed using the method described in
\cite[Ho et al. (2008)]{hoetal08}, and references therein, supplemented by
\cite[Potekhin \& Chabrier (2003)]{potekhinchabrier03}
for calculating Gaunt factors and
\cite[Suleimanov et al. (2012)]{suleimanovetal12}
to account for thermal effects
(see also \cite[Pavlov \& Panov 1976]{pavlovpanov76};
\cite[Potekhin 2010]{potekhin10};
\cite[Suleimanov et al. 2010]{suleimanovetal10}).
Examples of these spectra are shown in \cite[Ho (2013)]{ho13}.
Note that these weak magnetic field spectra assume a fully ionized hydrogen
atmosphere; partially ionized spectra are the subject of current work.

\begin{figure}[tb]
\begin{center}
 \includegraphics[width=3.4in]{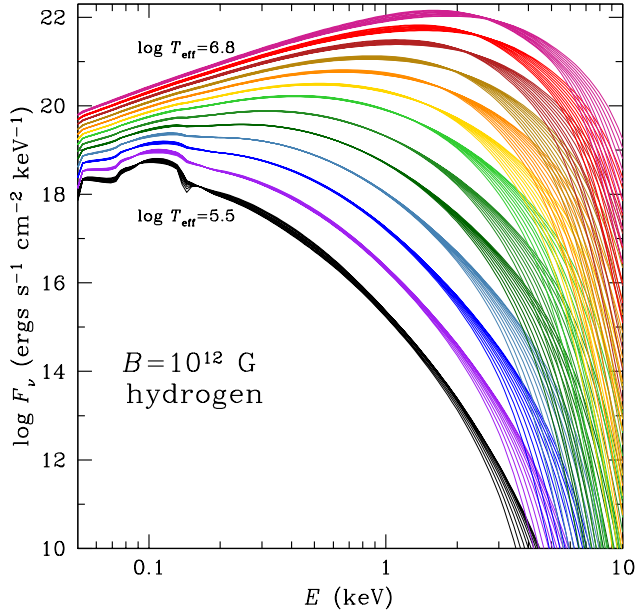}
 \caption{
Partially ionized hydrogen atmosphere model spectra for
effective temperatures $\log T_{\mathrm{eff}}=5.5-6.7$,
surface gravities $\log g=13.6-15.4$, and
magnetic field $B=10^{12}\mbox{ G}$.
}
   \label{fig:sp12}
\end{center}
\end{figure}

\begin{figure}[tb]
\begin{center}
 \includegraphics[width=3.4in]{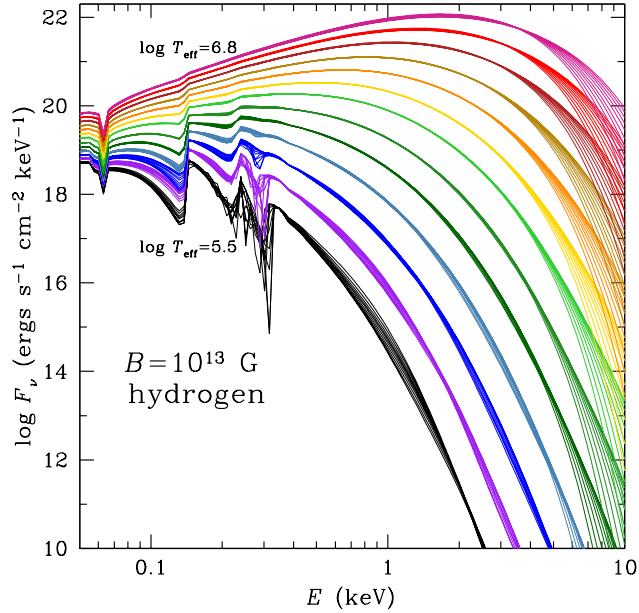}
 \caption{
Partially ionized hydrogen atmosphere model spectra for
effective temperatures $\log T_{\mathrm{eff}}=5.5-6.7$,
surface gravities $\log g=13.6-15.4$, and
magnetic field $B=10^{13}\mbox{ G}$.
}
   \label{fig:sp13}
\end{center}
\end{figure}

The second important difference between NSMAXG and NSMAX is a change in
XSPEC fit parameters.
Most XSPEC neutron star atmosphere models (NSA, NSAGRAV, and NSATMOS)
use the fit parameters  $T_{\mathrm{eff}}$, $M$, $R$, and either distance
$d$ or flux normalization $A$.
The last two are equivalent since $A\propto R^2/d^2$.
On the other hand, the fit parameters of NSMAX are  $T_{\mathrm{eff}}$, $A$,
and gravitational redshift $1+z_{\mathrm{g}}$ [$=(1-2GM/c^2R)^{-1/2}$].
There are two reasons for this different choice for NSMAX.
The first is that {\it all} models in XSPEC are calculated assuming
that emission arises from the entire visible surface of the neutron star,
i.e., $R=R_{\mathrm{NS}}$.  Thus the same value of $R$ must be used
to calculate $z_{\mathrm{g}}$.
The second reason is that NSMAX is constructed for particular values of
surface gravity $g$
[$=(1+z_{\mathrm{g}})GM/R^2$] (similarly NSA is calculated using a single
value of $g$, as well as assuming a fully ionized plasma, which is in
contrast to the partially ionized plasma of NSMAX).
Thus allowing $M$ and $R$ to vary as fit parameters would not produce
consistent results, i.e., the derived values of $M$ and $R$ would not
necessarily correspond to the value of $g$ that is used to compute the
atmosphere spectrum
(see \cite[Heinke et al. 2006]{heinkeetal06}, for comparisons between
NSA and NSATMOS).
We rectify this second issue by calculating spectra for a range of surface
gravities, i.e., $\log g\mbox{(cm s$^{-2}$)}=13.6-15.4$, thus allowing
NSMAXG to use $M$ and $R$ as consistent fit parameters.
Note that NSATMOS and NSAGRAV are also calculated for a range of $g$;
however recall that these two are non-magnetic models.
Figures~\ref{fig:sp12} and \ref{fig:sp13} show the resulting NSMAXG
spectra for $B=10^{12}$ and $10^{13}\mbox{ G}$, respectively.
Incidentally, the spectral tables of NSMAX and NSMAXG can easily be made
compatible for use with the other XSPEC neutron star fitting routines.

The spectra shown in Figs.~\ref{fig:sp12} and \ref{fig:sp13} only describe
emission from either a local patch of the stellar surface with a particular
effective temperature and magnetic field or a star with a uniform temperature
and radial magnetic field of uniform strength.  By taking into account
surface magnetic field and temperature distributions, we can construct
more physical models of neutron star emission,
which can be used for interpreting and decoding observations
(see \cite[Ho 2007]{ho07}; \cite[Ng et al. 2012]{ngetal12}, for details
and examples).

~\\
WCGH is grateful to his collaborators, Gilles Chabrier, Kaya Mori, and
Alexander Potekhin.
WCGH appreciates the use of the computer facilities at KIPAC
and acknowledges support from the IAU and STFC in the UK.

\end{document}